\documentclass{aa}
\usepackage{apj2aa}
\usepackage{astron} 
\usepackage{float,epsfig,psfig}
\usepackage{pstricks}

\def\hea4{{\it HEAO~A4}}
\def\heaoa2{{\it HEAO~A2}}
\def\heao1{{\it HEAO~1}}

\def\h0{$H_{\rm o}=50$~km~s$^{-1}$~Mpc$^{-1}$}
\def\q0{$q_{\rm o}$}

\def\etal    {{et~al.}~}

\def\cms3  {~{cm$^{-3}$}}

\newcommand{\mincir}{\raise
  -2.truept\hbox{\rlap{\hbox{$\sim$}}\raise5.truept \hbox{$<$}\ }}
\newcommand{\magcir}{\raise
  -2.truept\hbox{\rlap{\hbox{$\sim$}}\raise5.truept \hbox{$>$}\ }}

\begin{document}

\title{XMM-Newton Witness of M86 X-ray Metamorphosis} 
\author{A. Finoguenov\inst{1,3}, W. Pietsch\inst{1},
B. Aschenbach\inst{1}, F. Miniati\inst{2}}

\institute{Max-Planck Institut f\"ur extraterrestrische Physik,
             Giessenbachstra\ss e 1, D-85748 Garching, Germany 
\and
Max-Planck Institut f\"ur Astrophysik, Karl-Schwarzschild Str 1, 
D-85748 Garching,  Germany
\and
Space Research Institute, Profsoyuznaya 84/32, Moscow, 117810, Russia
}

\date{Received 2003, May 28; accepted 2003, October 16}
\authorrunning{Finoguenov \etal}

\abstract{The environmental influence of cluster media on its member
galaxies, known as Butcher--Oemler effect, has recently been subject to
revision due to numerous observations of strong morphological
transformations occurring outside the cluster virial radii, caused by some
unidentified gas removal processes. In this context we present new
XMM-Newton observations of M86 group.  The unique combination of high
spatial and spectral resolution and large field of view of XMM-Newton allows
an in-depth investigation of the processes involved in the spectacular
disruption of this object.  We identify a possible shock with Mach number of
$\sim 1.4$ in the process of crushing the galaxy in the North-East
direction. The latter is ascribed to the presence of a dense X-ray emitting
filament, previously revealed in the RASS data. The shock is not associated
with other previously identified features of M86 X-ray emission, such as the
plume, the north-eastern arm and the southern extension, which are found to
have low entropy, similar to the inner 2 kpc of M86.  Finally, mere
existence of the large scale gas halo around the M86 group, suggests that
the disruptions of M86's X-ray halo may be caused by small-scale types of
interactions such as galaxy-galaxy collisions.  }

\maketitle

\section{Introduction}

X-ray studies of galaxies in nearby clusters provide important clues on
environmental effects on the properties of galaxies. X-ray luminous
early-type galaxies appear often surrounded by group-size mass
concentrations, whose interaction with the cluster environment is not yet
completely understood from both observational and modeling sides. In fact,
on one hand, recent observations find quite a bit of substructure in merging
systems, demonstrating the survival of the core of the accreting objects
(e.g. Vikhlinin et al. 2001).  On the other hand, detailed studies of
cluster galaxies show that strong interactions take place among the
infalling galaxies at the virial radius (Poggianti et al. 1999; Goto et
al. 2003). In addition, recent studies on the evolution of galaxies prior to
their infall onto clusters reveal strong effects of local environment on the
appearance of galaxies (Kodama et al. 2001). Therefore, the morphology of
galaxies appear to be disturbed in the so called quiet accretion mode of
cluster formation.

M86 is a prototype for this type of investigation. It is a giant elliptical
galaxy falling at high speed toward M87 (relative velocity $\sim 1500$ km
s$^{-1}$). Furthermore it sits in the core of a larger structure, M86 groups
of galaxies, which exhibits a diffuse X-ray emission on a few hundreds kpc
scale (B\"ohringer et al. 1994). An X-ray study of M86 with Einstein (Forman
et al. 1979) revealed a peak of emission centered on M86 and a plume
extending northwest of the galaxy. Ram-pressure stripping of the hot gas
from M86 as the galaxy traverses the Virgo cluster was suggested to explain
the observed structure (Forman et al. 1979; Fabian, Schwartz, \& Forman
1980; Takeda, Nulsen, \& Fabian 1984).  Spatial analysis of the X-ray
brightness of M86 from the ROSAT PSPC and HRI data was presented in
Rangarajan et al. (1995). In addition to the plume, they identified a
southern extension (extending from the center by about $3^\prime$ to the
south), a void (north to the galaxy), and a northeastern arm (extending from
the center by more than $5^\prime$ to the northeast). Detailed spectral
analysis of M86 emission using ROSAT PSPC and ASCA SIS results was presented
in Finoguenov \& Jones (2000). The temperature variations were found to lie
in the 0.6--1.1 keV range. Cool regions include the galaxy center and extend
eastward with slightly increasing temperature. In the plume, spectral
variations were detected on the 0.1 keV level.

\begin{figure*}
\includegraphics[width=5.9cm]{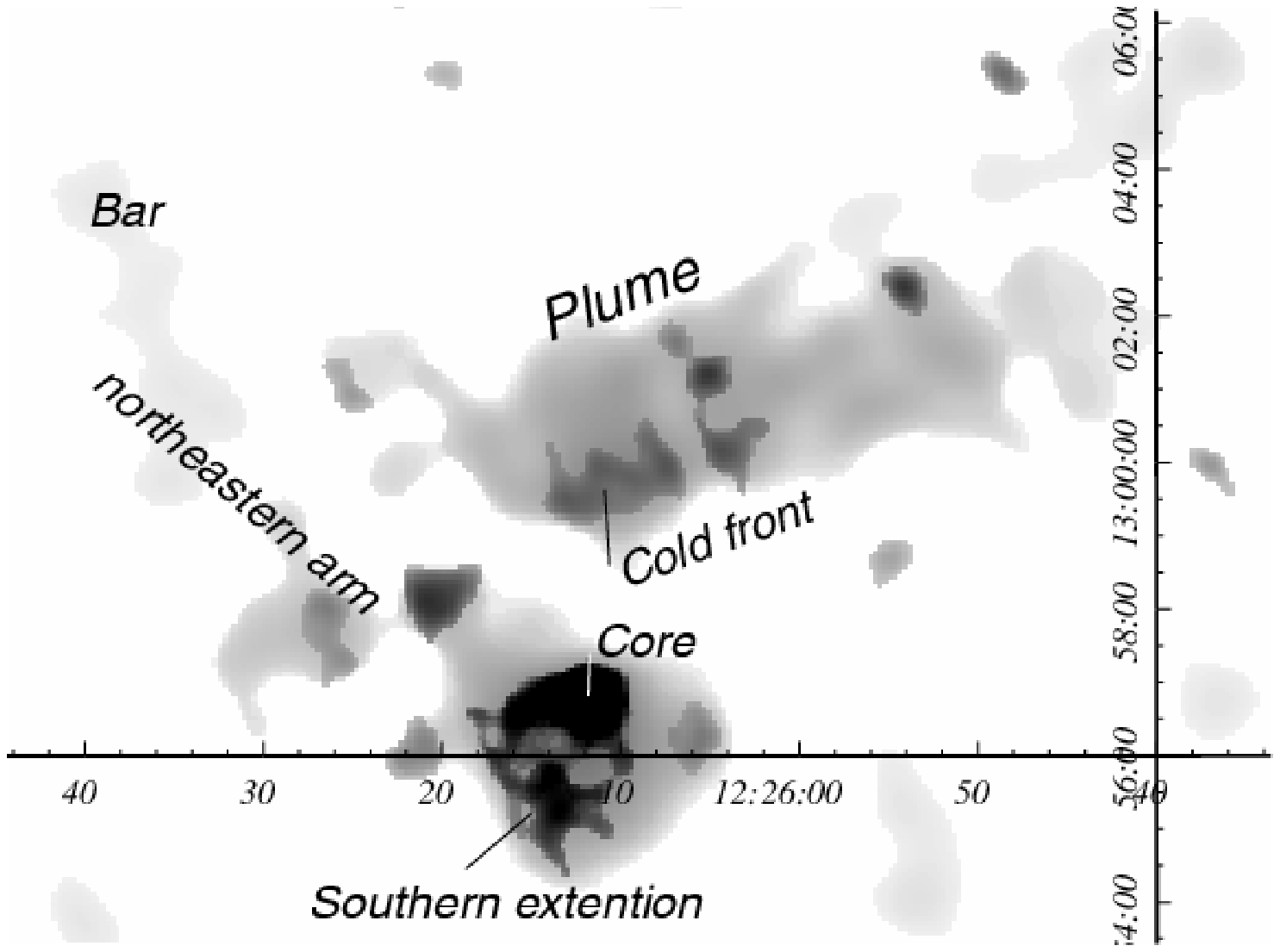}\includegraphics[width=5.9cm]{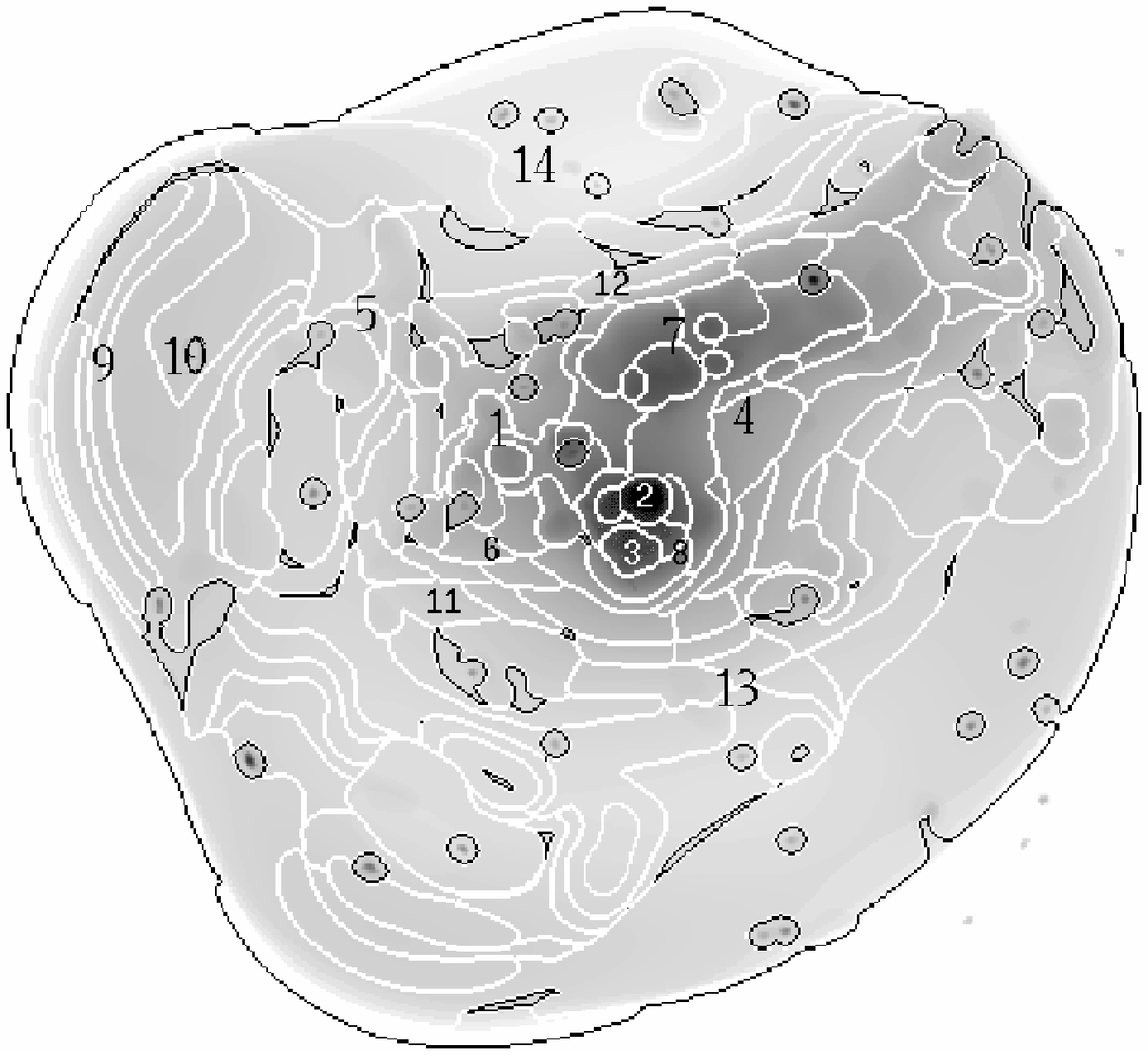}\includegraphics[width=5.9cm]{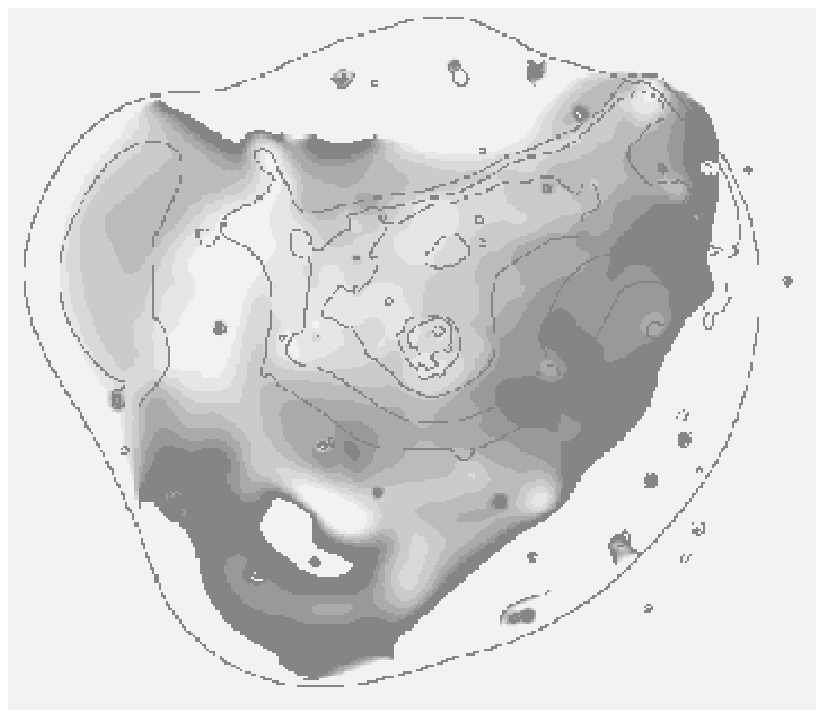}

\figcaption{{\it Left panel.} Identification of the previously reported
features of M86 on the smallest scale of the wavelet decomposition of the
image in the 0.5--2 keV band. {\it Middle panel.} Wavelet-decomposed image
of M86 with contours indicating the selection of regions for the spectral
analysis. Black contours denote the regions excluded from the analysis due
to either low statistics or a presence of the point source. Numbers on the
figure indicate the approximate location of the larger zones, reported in
Tab.\ref{t:mo}. {\it Right panel.} Hardness ratio map with surface
brightness contours overlaid. Dark grey color corresponds to temperature of
1.1--1.2 keV, grey to 0.9--1 keV, light grey to 0.7--0.9 keV, white to
0.5--0.7 keV.
\label{imh}
}
\end{figure*}

As it became evident from the temperature and detailed surface
brightness maps, M86 does not fit into the simple picture suggested by
early observations.  For this reason Rangarajan et al. (1995) proposed
that M86 owes its appearance to the presence of a conical shock seen
in projection and due to the galaxy interaction with Virgo
intracluster medium (ICM).

In this paper we present new XMM-Newton observations of M86 galaxy. By
taking advantage of the unique combination of high spatial and spectral
resolution of XMM-Newton, we study the thermodynamic properties of a total
of 120 regions comprising the environment in and around M86.  Based on
these, we were able to reveal the remarkable presence a of a weak shock in
the process of crushing the galaxy interstellar medium (ISM).  In view of a
$2.4\pm 1.4$ Mpc distance of M86 to the core of Virgo (Neilsen \& Tsvetanov
2000), it is, however, quite unlikely for such shock to be driven by the
impact of the galaxy on Virgo ICM, even considering the uncertainties in the
distance determination. A conclusion that M86 group is located outside the
M87 cloud is further supported by observation of the X-ray halo extending
over a degree in diameter (B\"ohringer et al. 1994).  We have therefore
carried out a closer inspection of the large scale gas distribution
surrounding M86 which led us to associate the counteracting structure with a
filament in the neighborhood of Virgo cluster, already discovered by
B\"ohringer et al. (1994).

The picture of M86 system emerging from these observations argues for
a series of interactions at different stages of completeness.
Although the weak shock is probably driven by an interaction with the
surrounding large scale medium, the large-scale emission of M86 group
is mostly unchanged. On the contrary, the core of M86 galaxy has been
already strongly perturbed. This suggests that additional significant
dynamic interactions were responsible for disturbing the X-ray
morphology of M86 galaxy and that they took place on small scales, of
order 10--50 kpc. Galaxy-galaxy encounters may offer a possible
explanation.

The paper is organized as follows: in \S\ref{s:data} we present
a detailed description of the data, including the reconstruction of
maps of thermodynamic quantities in \S\ref{s:maps} that lead to
the identification of a weak shock in \S\ref{s:shock}. The results 
are discussed in \S\ref{s:disc} and summarized in \S\ref{s:conc}.

In the following we will assume a 19.5 Mpc distance to M86 (Neilsen \&
Tsvetanov 2000), for which $1^\prime=5.67$ kpc.

\section{Observations and data reduction}\label{s:data}

M86 was observed by XMM-Newton (Jansen et al 2001) during July 1--2 2002, as
a part of the GTO program of the telescope scientists at MPE. Screening of
the observations to remove particle background flaring episodes results in
the net exposure time of 46 ksec, 64 ksec and 67 ksec for pn, MOS1 and MOS2
EPIC detectors. pn observations were performed with medium filter and in the
extended full frame observing mode (Str\"uder et al. 2001), which is
characterized by a reduced level of out-of-time events. MOS1 and MOS2
(Turner et al. 2001) observations were performed with the thick filter and
in the present study are used only for the imaging analysis. The advantage
of using MOS and RGS spectra is mostly in the detailed element abundance
study (particularly given the thick filter choice for the MOS and a typical
temperature of the X-ray emission of M86), which will be given in a separate
paper.

Initial steps of data reduction were performed using XMMSAS 5.4. The
spectral analysis of pn data was carried out in the 0.4--5 keV band. For the
background subtraction we used several background accumulations, one by Read
(2003) and the other performed close to the observation
(e.g. APM08279+5255; Hubble Deep Field South). We find that for the M86
observation, the detector background in the 10--15 keV range corresponds
better to observations performed at similar epochs. Although the details of
the background subtraction are not very critical for the analysis of the
soft emission of M86, it could be of some importance for the analysis of the
hard component, associated with unresolved LMXBs.

The vignetting correction is performed taking the source extent and a recent
vignetting calibration (Lumb et al. 2002b) into account, which is mostly
important for the absolute flux determination, given our choice of the
energy range. Remaining systematic uncertainty of the flux is below 4\% for
both pn and MOS (Lumb et al 2002b). 

\begin{figure*}
\includegraphics[width=17.cm]{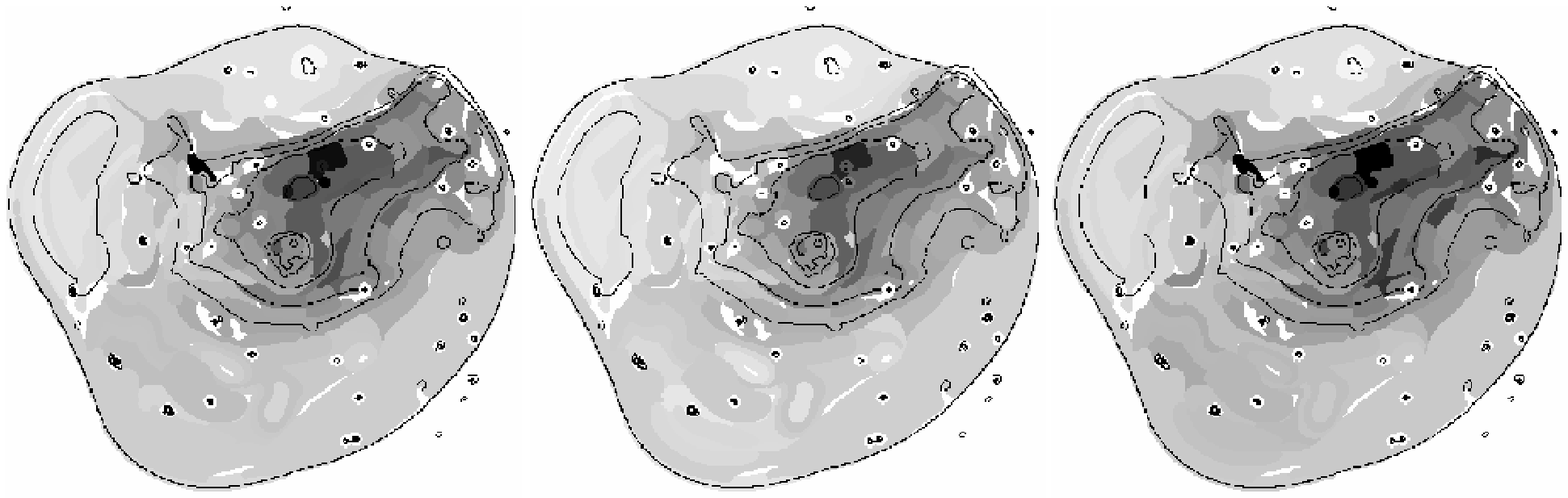} 
\figcaption{Fe abundance map, overlaid with surface brightness contours.
  From left to right the panels show the best-fit values, low and
  upper limit on Fe abundance, ranging from 0 (light grey) to 0.7
  (black) of solar.
\label{fe}}
\includegraphics[width=17.cm]{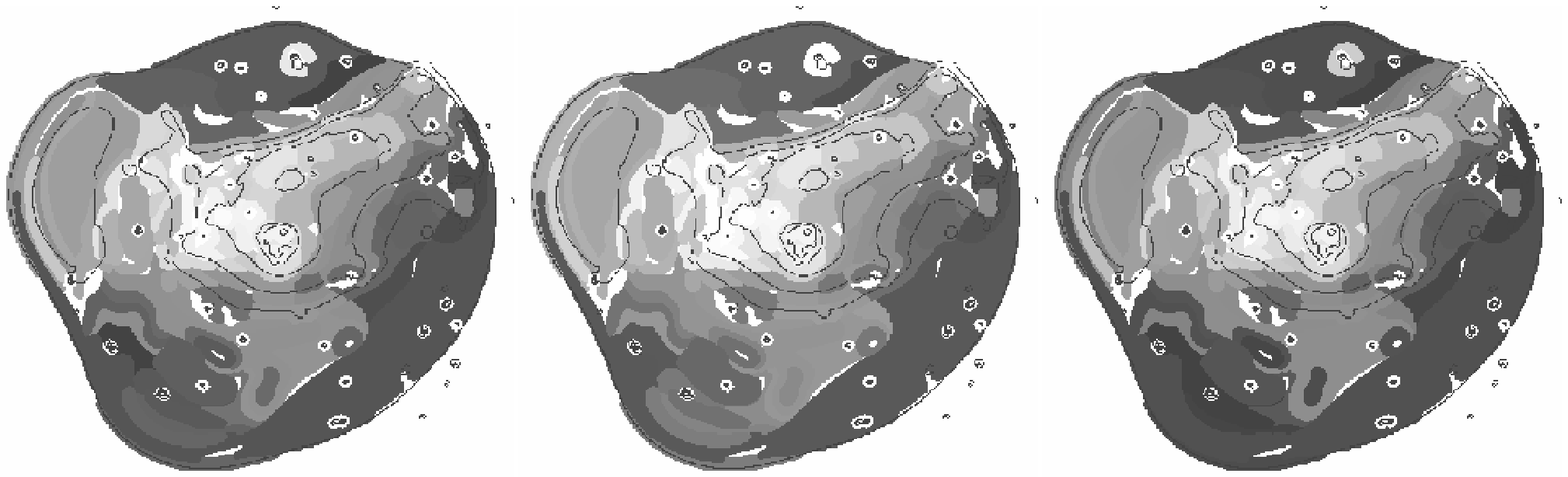} 
\figcaption{Temperature
map, overlaid with surface brightness contours. From left to right the
panels show the best-fit values, low and upper limit on temperature,
ranging from 0.6 (light grey) to 1.2 (dark grey) keV.
\label{kt}}

\includegraphics[width=17.cm]{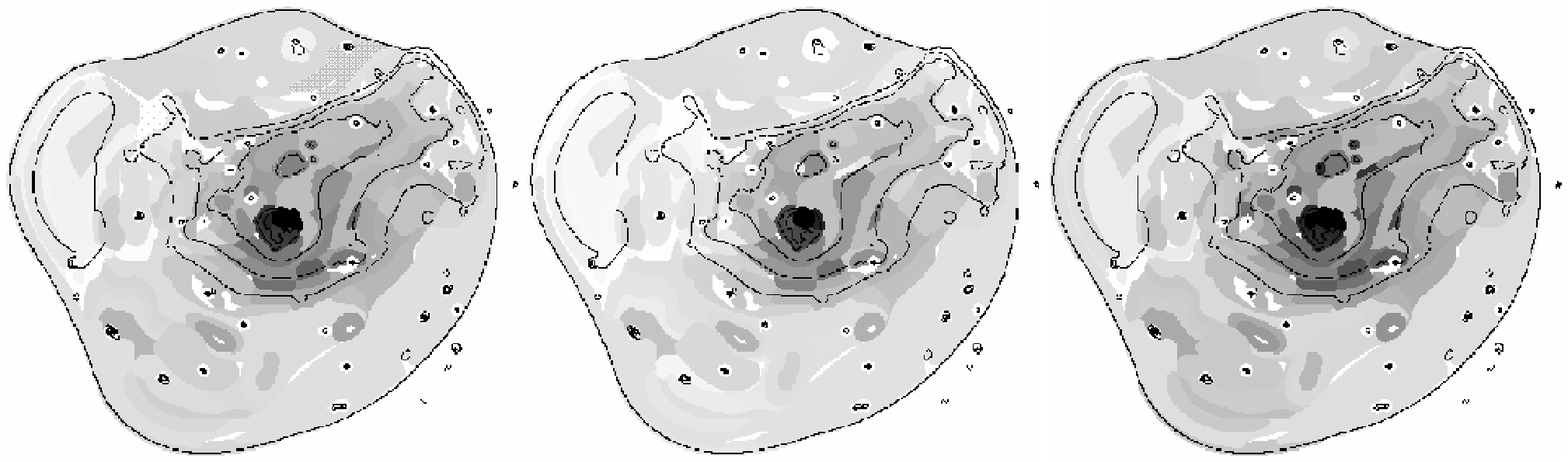} 
\figcaption{Pressure
map, overlaid with surface brightness contours. From left to right the
panels show the best-fit values, low and upper limit on pressure,
ranging from $2\times10^{-12}$ ergs cm$^{-3}$ (light grey) to
$2\times10^{-11}$ ergs cm$^{-3}$ (black).
\label{f:p}}
\includegraphics[width=17.cm]{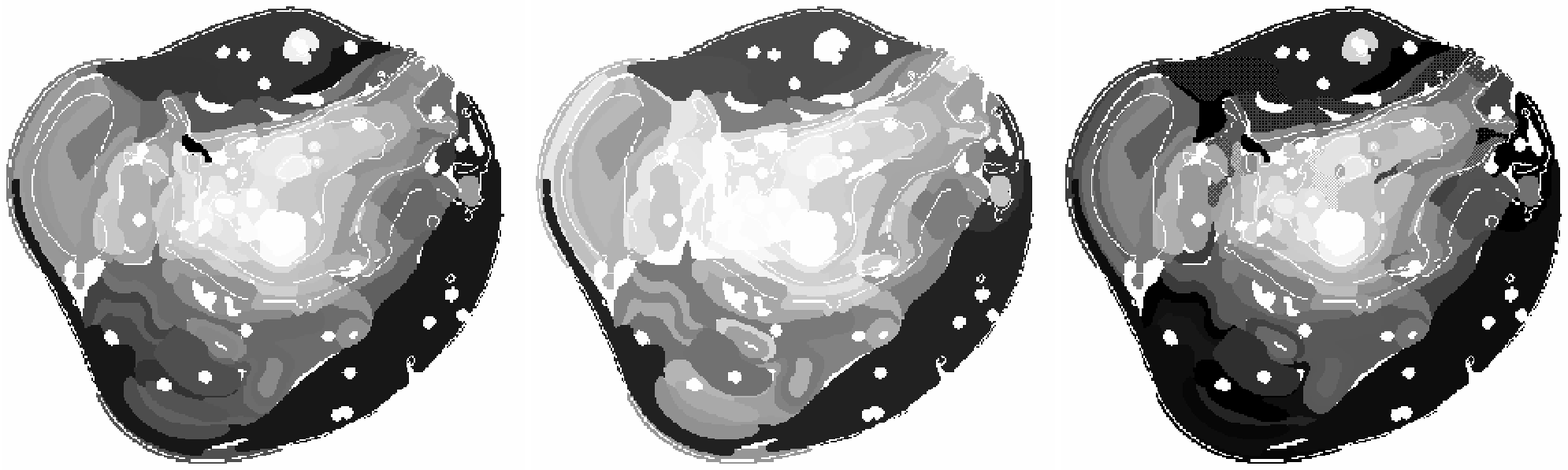} 
\figcaption{Entropy map, overlaid with surface brightness contours. 
From left to right the panels show the best-fit
  values, low and upper limit on entropy, ranging from 8 keV
  cm$^2$ (white) to 100 keV cm$^2$ (black).
\label{f:s}}
\end{figure*}

In selecting the regions for the spectral analysis we use a combination of
the surface brightness and hardness ratio maps to emphasize regions of
similar color and intensity. Fig.\ref{imh} shows both maps, where the
surface brightness is a wavelet-reconstructed (Vikhlinin et al. 1995) image
in the 0.5--2 keV band and the hardness of the emission is a ratio of the
wavelet-reconstructed images in the 0.5--1 and 1--2 keV bands. An advantage
of using wavelets consists in background removal by spatial filtering and a
control over the statistical significance of the detected
structure. Complications arise due to splitting the image into discrete
scales, which we overcome by additional smoothing applied before producing
the hardness ratio map. More details can be found in Briel et al. (2003).
In combining MOS and pn, we differentially normalize the exposure maps to
account for the difference in the sensitivity of MOS vs pn to the
temperatures in the 0.6--1.2 keV range, given our selection of bands. In
doing so, we avoid variations in color due to different sky coverage
provided by MOS and pn. Given the choice of the energy bands, variation in
the Fe abundance should not play a strong role in the color selection,
however iron line emission determines up to 90\% of the flux in the 0.5--2
keV band. To establish the mask file for the further spectral analysis we
use the changes in the hardness ratio that correspond to temperature in the
ranges 0.48--0.56--0.60--0.64--0.72--0.8--0.9--1.0--1.1--1.2 keV and have
equal intensity within a factor of two. Taking the isolated regions of equal
color and intensity separately and imposing the criterion that the regions
should be larger than the PSF ($15^{\prime\prime}$) width and contain more
than 1000 counts in the raw pn image, we obtain the final mask used in the
spectral extraction, which we display as contours in the central panel of
Fig.\ref{imh}.

For the spectral fits, we first assume a single temperature plasma in
collisional ionization equilibrium, with emission due to thermal
bremsstrahlung dominated by H, He, C and O ions and line emission which, at
the temperature typical of M86 and in the energy range of XMM-Newton, is
dominated by Fe (e.g. Raymond \& Smith 1975).  For the purpose we use the
recent revision of the initial Raymond--Smith code, APEC (Smith et
al. 2001), implemented in the XSPEC v.11.2, which benefits from improved
treatment of the Fe L-shell line atomic physics (e.g. Liedahl et al. 1995).
In the spectral analysis, we leave temperature and abundances of O, Ne, Mg,
Si, Fe as free parameters. C abundance, which cannot be fit, but contributes
to the continuum emission is set to 0.2 times the solar value. Setting C/O
ratio to solar gives the same results.

As a second step, a power law component which, according to several studies
well describes the emission from low mass X-ray binaries (LMXB) (Finoguenov
\& Jones 2001, 2002; Irwin et al. 2003) was introduced. We need to do so,
owing to high level of diffuse X-ray emission of M86 and the PSF of the XMM
telescopes, which prevent us from resolving LMXB individually. The spectral
index found by fitting the regions close to the M86 core and is 1.5,
consistent with the results for other galaxies (Irwin et al. 2003). In the
following we fix the slope and only fit the normalization of the model.

Finally few regions close to the center of M86 still require a more complex
spectral model, which we achieve by adding a second plasma (APEC)
component. Since the emission weighted temperature has not changed within
estimated statistical errors, we conclude that the results from the previous
steps on temperature structure are still valid. In all the fits $N_H$ was
fixed at the Galactic value of $2.7\times10^{20}$ cm$^{-2}$ (Stark et
al. 1992).

\begin{figure*}

\includegraphics[width=17cm]{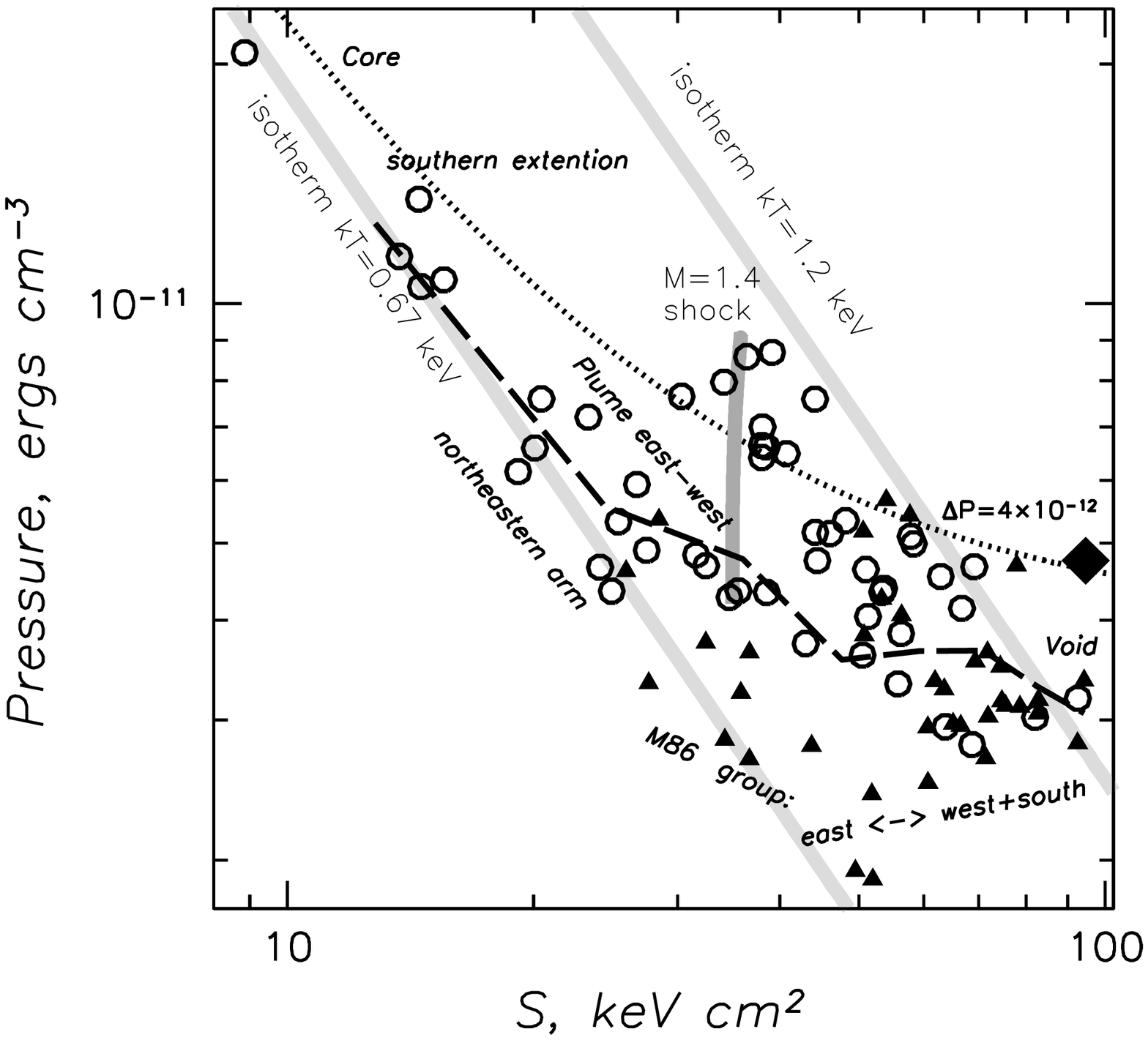}

\figcaption{Pressure vs entropy for all the regions in M86. Only the points
with significance greater than 4 sigma are shown. Open circles and filled
triangles denote the points with high and low Fe abundance,
respectively. Light gray lines show isotherms and a dark-gray line shows
the shock adiabat of Mach number of 1.4. Dotted line shows the prediction
on imposing an additional pressure of $4\times10^{-12}$ ergs cm$^{-3}$ to
an initially isothermal gas with $kT=0.67$ keV. Filled diamond displays the
characteristics of the projected component found in the east to M86.
\label{p_vs_s}
}

\end{figure*}

\subsection{Maps}\label{s:maps}

The spectral analysis provides statistically significant data on kT for all
of the 120 selected regions.  Metal abundances and normalizations are well
constrained for Fe in most of those regions, for Mg and Si for some of them,
and for O and Ne in only a few of them. In Fig.\ref{fe}--\ref{kt} we show
the temperature and iron abundance maps. We find temperature variations in
the 0.6--1.2 keV range, in general agreement with the previous findings. In
addition several new details are revealed by these maps: the northeastern
arm, for example, is clearly seen as a part of the cold spiraling thread
starting from the center and ending at $5^\prime$ towards the north-east.
More details variations are seen in the ``plume'', making it progressively
hotter towards north-west. In the Fe abundance map, two zones are clearly
distinct. There is an outer zone with Fe abundance of 0.1--0.2 solar and an
inner zone, with Fe abundance in excess of 0.3 solar, the latter reaching
0.7 solar in some zones in the plume.

To estimate the pressure and entropy in each region, we need to estimate the
length of the column for each selected two-dimensional region on the sky. We
assumed a geometry, in which every region is part of a spherical shell that
is centered on the core of M86 system and has its inner and outer radii
passing through the nearest and furthest points of the selected region,
respectively. This spherical shell is further intersected by a cylinder,
that is directed towards the observer and in the observer plane has the
cross-section of the selected region. For the concentric regions our
geometry corresponds to a usual ``onion peeling'' technique
(e.g. Finoguenov, Ponman 1999). The advantage of our assumption is that the
projected size depends only on the distance of the region to the center of
the galaxy and not on the size of the extraction region. Alternatively, one
could assume a projected dimention of the structure to be some combination
of the observed dimentions. For large identified features, such as plume,
north-eastern arm, and the shock, discussed below this makes less than 20\%
difference. Smaller features of such a geometry are unlikely to be detected,
unless they correspond to an embedded low-entropy gas. For these we
underestimate the pressure and overestimate the entropy by typically a
factor of 1.3. Using the adopted line-of-sight estimation method, in
Figs.\ref{f:p}--\ref{f:s} we present the derived pressure and entropy
maps. Note that we do not attempt to account for the effect outer components
can have on the derived parameters of the inner components through the
geometrical projection. However, projection is known to have a mild effect
in X-ray systems with $\beta>0.5$, as is the case southwards from M86
(Finoguenov \& Jones 2000), where we find the shock.

Low entropy zones include the core, part of the plume close to the core and
the northeastern arm. Highest entropy regions form a circle surrounding
M86 system, suggesting that the large-scale emission of M86 group is
unperturbed. The pressure map also shows that the void region has a pressure
typical of M86 group, in agreement with a conclusion of Rangarajan et
al. (1995), and it is rather M86-galaxy that exhibits large contrast in the
pressure map, which we analyze below.

\subsection{Revealing the shock}\label{s:shock}

In Fig.\ref{p_vs_s} we plot the pressure ($ n kT$) vs entropy ($kT n^{-{2
\over 3}}$) for all the regions for which the statistical errors on the
pressure and entropy are less than 25\%.  We assign the data to either
M86-group (filled triangles) or M86-galaxy (open circles), depending on
whether the Fe abundance is above or below 1/4 solar, respectively. This
threshold is higher than a typical element abundance for groups of 0.1 solar
(e.g. Finoguenov et al. 2002), but allows a more robust criterion for
intragroup media (IGM)--ISM separation given the typical error on the iron
abundance of 0.1 solar. This method is independent of geometry assumptions
and it allows us to check whether or not such assumptions affect our
derivation of thermodynamic quantities. We checked that using other elements
leads to similar results, where statistics allows it.

\includegraphics[width=8.4cm]{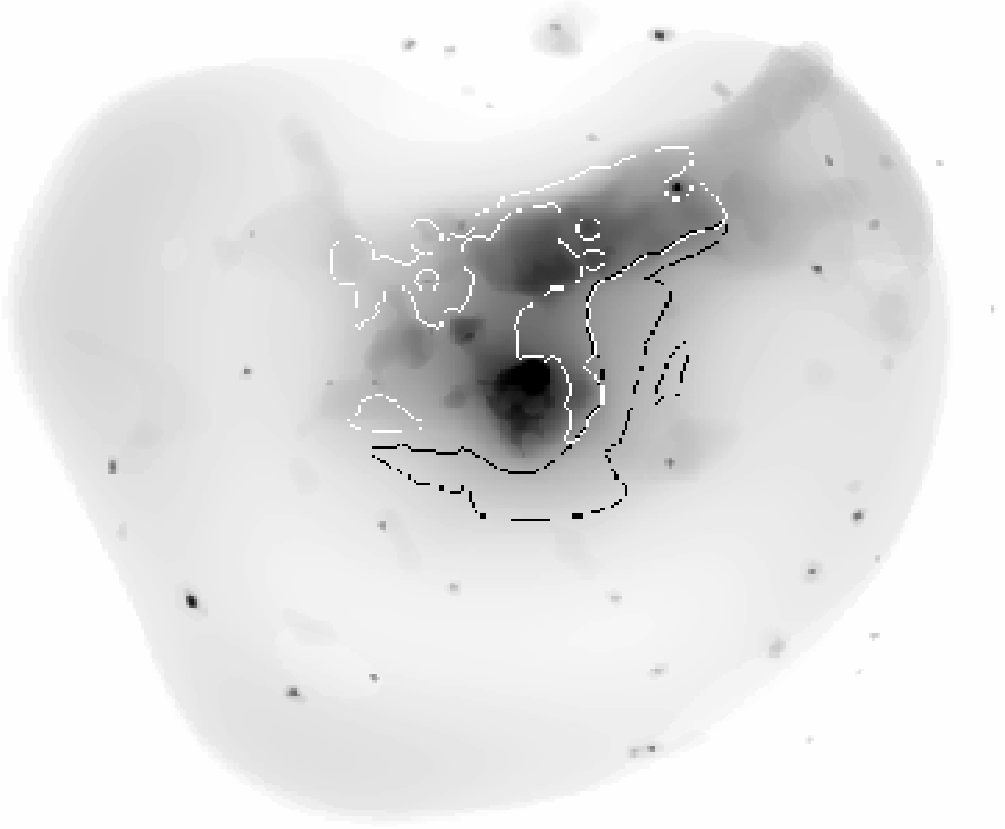}

\figcaption{Image of M86. Black contours indicate shocked region, while
  white contours indicate the preshock regions of the same entropy and
  similarly high metallicity.
\label{imh_shock}
}

In Fig.\ref{p_vs_s} we also show two isotherms, corresponding to
$kT=0.67$ and 1.2 keV. Any uncertainty due to assumptions of geometry
would move the data points only along isotherms. The high pressure,
low entropy core is located on the top left corner of the plot. The
plot shows that, as we move towards the outer regions of the galaxy
the pressure drops while the entropy increases. Up to a few tens keV
cm$^2$, the selected regions are well aligned along a dashed line
representing a mass-weighted mean of the data. We regard this behavior
as typical as it compares well with ROSAT/ASCA results on other
early-type galaxies from an extensive tabulation of Sanderson et
al. (2003). Between 30 and 40 keV cm$^2$ we notice that the data
points suddenly appear distributed along two main loci with a gap in
between. The origin of this could be ascribed either to a number of
reason such as enhanced data scatter and/or peculiar projection
effects. However it could have a more interesting physical origin,
specifically it could be due to the presence of a weak shock crossing
M86 medium. In fact the points along the two separate loci are well
connected by a Rankine-Hugoniot shock adiabat corresponding to a Mach
number shock of about 1.4.  It is worth pointing out that a weak shock
manifests primarily as a pressure jump and with only a minor entropy
change as ${\Delta S \over S} \propto ({\Delta P
\over P})^3$, as it is the case of Fig.\ref{p_vs_s}.

Additional support for our interpretation is provided by the spatial
distribution of the points connected by the shock adiabat. This is
shown in Fig.\ref{imh_shock} where we display the regions containing
the points with entropy between 30 and 40 keV cm$^2$.  Intriguingly,
the regions (diagram points) associated with each of the two loci in
the diagram, form two aligned, bow shaped and $180^\circ$ long strips,
with the low pressure region {\it closer} to M86 core. Such pressure
structure is clearly deviating from a hydrostatic profile. However, it
is naturally interpreted in the framework in which a reverse shock is
in the process of {\it crushing} M86 ISM.

As with any study of this kind, unavoidable projection effects will produce
some blending of pre- and post-shock regions reducing thus the estimate of
the shock Mach number.  A possible simplified geometry for the system M86
cloud - shock relative to an ideal observer is illustrated in
Fig.\ref{sch:shock}. In this depiction the shock is oblique with respect to
the line of sight. Only part of the shock is visible since most of the
emission from the region to the north-east of M86 is contributed by gas
associated with the preshock material.

The passage of even a weak shock increases the pressure of the gas crossing
through it. The shock weakens as is moves toward the inner, high pressure
core of M86 galaxy. In addition the pressure jump is lower in the outer
regions were the shock is oblique. Thus the region corresponding to an
entropy of 30-40 keV cm$^{-2}$ could represent the nose of the plunging gas
cloud, and the higher entropy ones the outer zones where the shock induced
pressure jumps are weaker.  A maximum pressure increase of $4\times10^{12}$
ergs cm$^{-3}$ is plotted in Fig.\ref{p_vs_s} to guide the eye.

\includegraphics[width=4.cm]{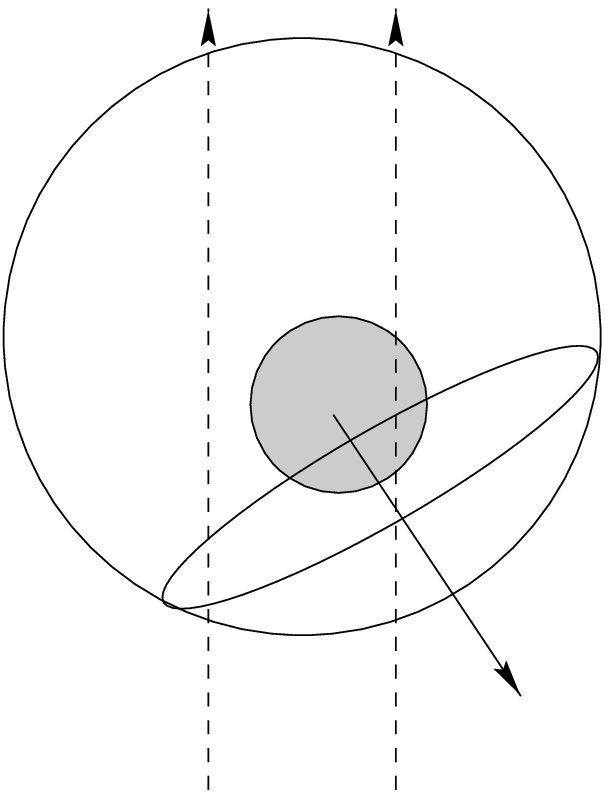}

\figcaption{Schematic view of the crossing of M86 by the plane of the shock
wave. Direction of motion of M86 galaxy relative to the observer is
indicated by a slight replacement of M86 galaxy (filled circle) relative to
the group (open circle).
\label{sch:shock}
}

\subsection{Additional Emission Components}

\includegraphics[width=8.4cm]{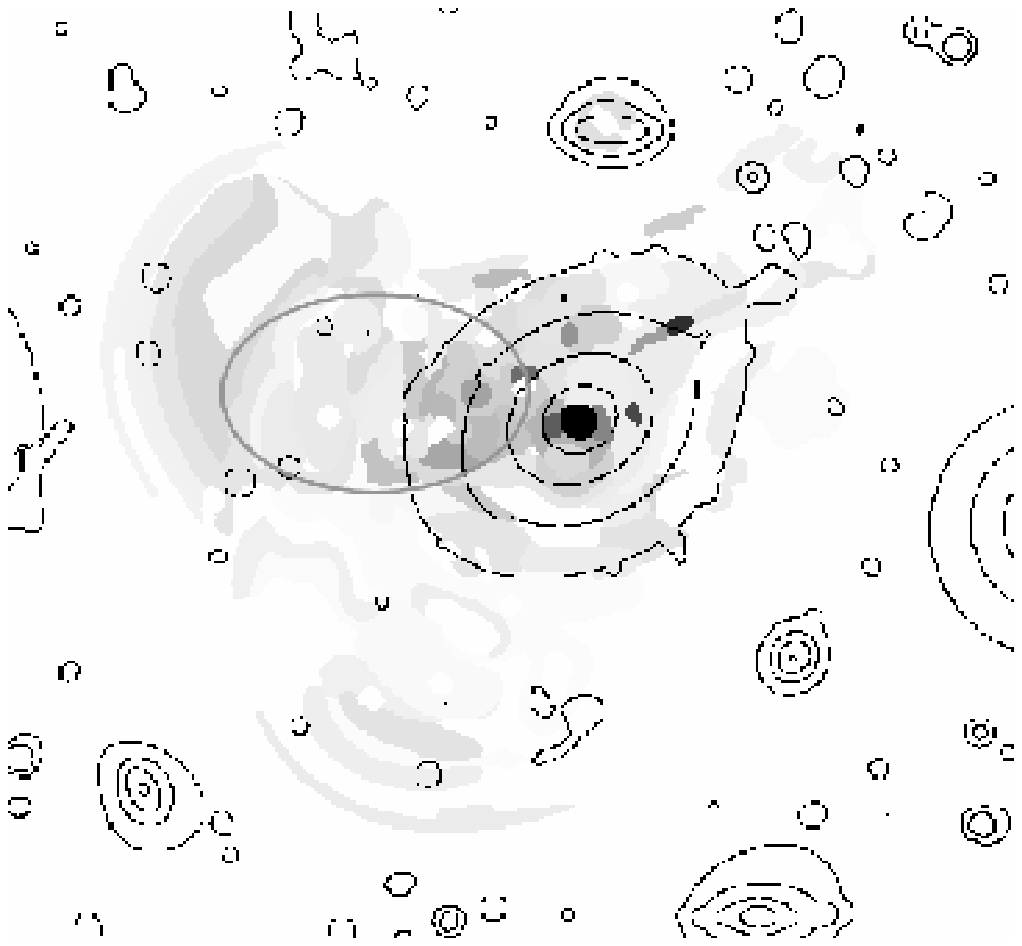}

\figcaption{Image of the intensity of the power law component per unit
extraction area in the X-ray spectra of M86.  Contours indicate the surface
brightness in the infrared, derived from DSS-2 data. Gray ellipse indicates
the region selected for a detailed spectral analysis, presented in
Fig.\ref{f:spe}. 
\label{f:hard}
}

In the previous section we presented the results for the main emission
component dominating the X-ray emission in the 0.5--2 keV band.  In addition
we have introduced an underlying power law component, assuming it originates
from unresolved point sources in M86.

As such, the spatial distribution of this
component should closely follow the distribution of the optical light of M86
(e.g. Finoguenov, Jones 2001). However, when we analyzed the distribution of
the intensity of the power law component, a strong excess in the northeast
was found, as illustrated in Fig.\ref{f:hard}. Thus, we have found a
disagreement with the assumption that the hard X-ray emission originates
only due to LMXB. We have extracted a spectrum from the large region in the
north-east to see in more detail the spectrum of the hard excess and
realized that it is of thermal origin, having a temperature of $1.5\pm0.1$
keV, as is shown in Fig.\ref{f:spe}.

\includegraphics[width=8.4cm]{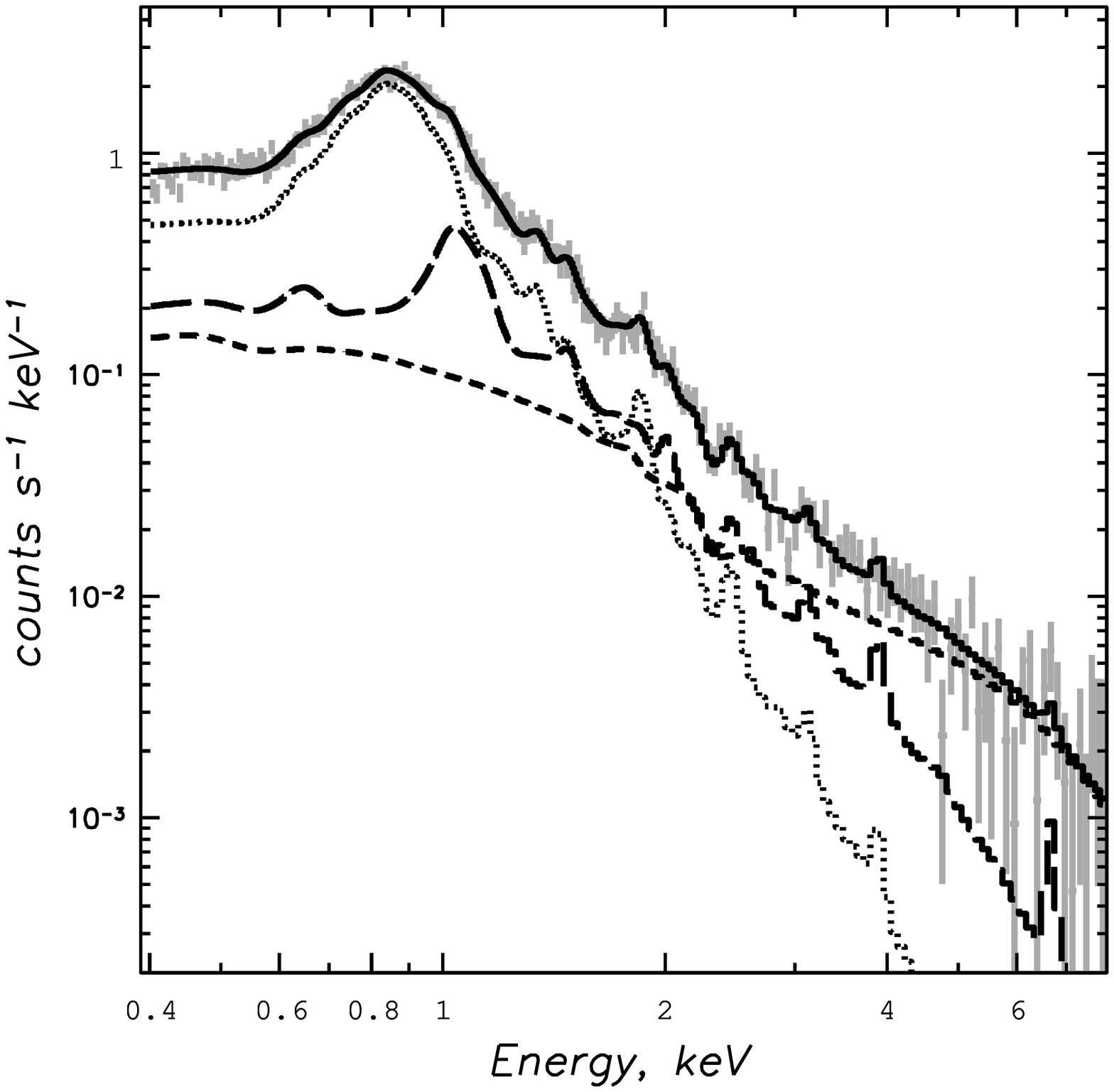}

\figcaption{Spectrum of the M86 zone with enhanced hard emission. Grey
  crosses indicate the data points, solid line shows the combined model,
  consisting of normal component (dotted), hotter component (long-dashed)
  and a power-law (short-dashed). 
\label{f:spe}
}

We have attempted an analysis with two temperature components, but the
statistics is not good enough to derive any safe conclusion for our choice
of the mask. In order to clearly separate this component one needs to group
together the zones of similar temperature. This is exactly what has been
done to extract this component. We believe this component corresponds to the
shocked gas of M86 seen in projection. Realistic placement of this component
into Fig.\ref{p_vs_s}, supports this scenario assigning this gas to M86
group. We have checked the spectra south-east and south-west, they all are
fitted by a 1.2 keV temperature plasma, ruling out a possibility of a dense
large-scale emission of 1.5 keV temperature in front of M86. Emission from
the Virgo cluster is not contributing strongly within the XMM FOV.

\subsection{Main properties of M86}

\begin{figure*}
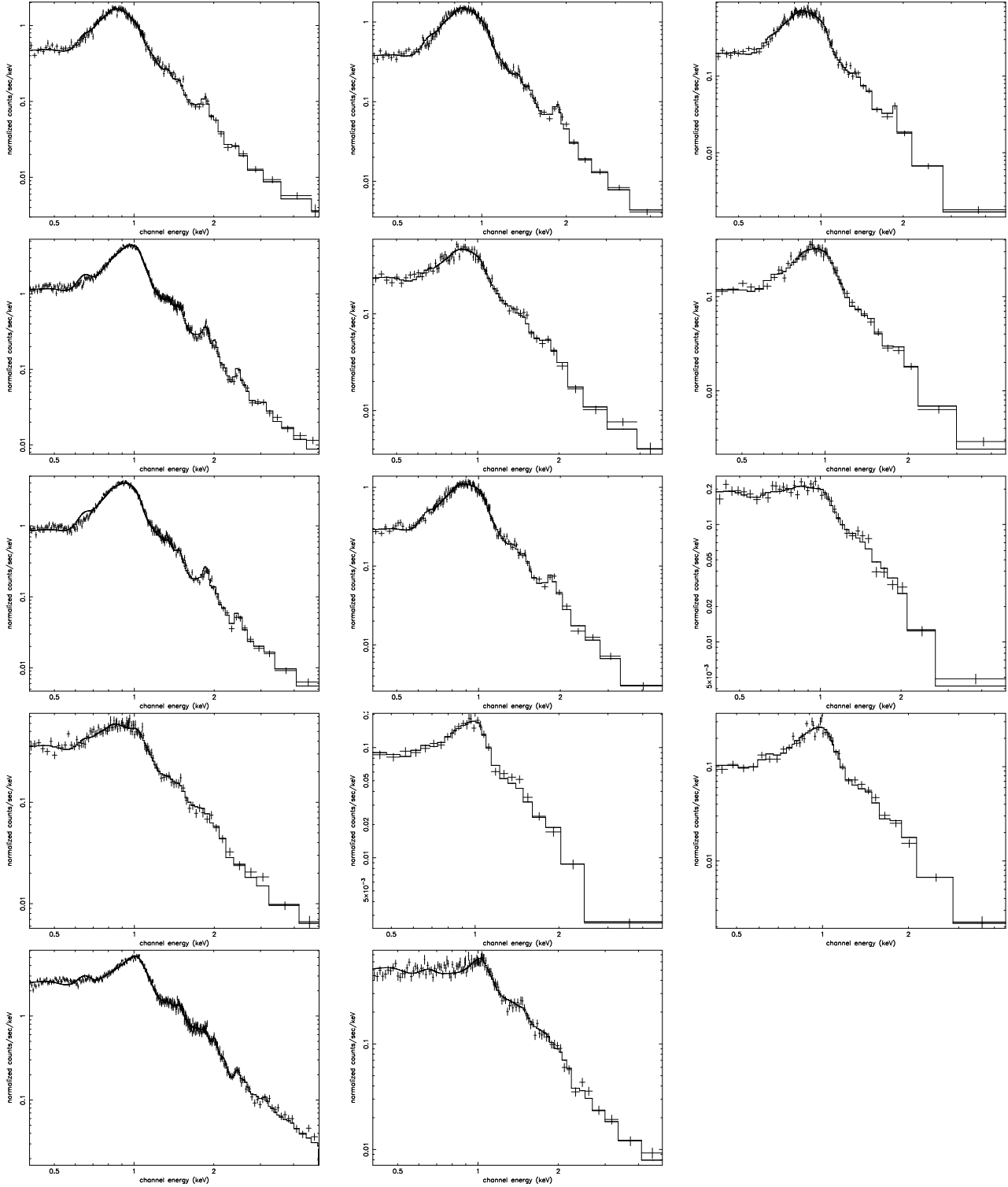

\includegraphics[bb=79 43 570 760,height=5.8cm,angle=-90,clip]{f8p01.ps}\includegraphics[bb=79 43 570 760,height=5.8cm,angle=-90,clip]{f8p02.ps}\includegraphics[bb=79 43 570 760,height=5.8cm,angle=-90,clip]{f8p03.ps}

\includegraphics[bb=79 43 570 760,height=5.8cm,angle=-90,clip]{f8p04.ps}\includegraphics[bb=79 43 570 760,height=5.8cm,angle=-90,clip]{f8p05.ps}\includegraphics[bb=79 43 570 760,height=5.8cm,angle=-90,clip]{f8p06.ps}

\includegraphics[bb=79 43 570 760,height=5.8cm,angle=-90,clip]{f8p07.ps}\includegraphics[bb=79 43 570 760,height=5.8cm,angle=-90,clip]{f8p08.ps}\includegraphics[bb=79 43 570 760,height=5.8cm,angle=-90,clip]{f8p09.ps}

\includegraphics[bb=79 43 570 760,height=5.8cm,angle=-90,clip]{f8p10.ps}\includegraphics[bb=79 43 570 760,height=5.8cm,angle=-90,clip]{f8p11.ps}\includegraphics[bb=79 43 570 760,height=5.8cm,angle=-90,clip]{f8p12.ps}

\includegraphics[bb=79 43 570 760,height=5.8cm,angle=-90,clip]{f8p13.ps}\includegraphics[bb=79 43 570 760,height=5.8cm,angle=-90,clip]{f8p14.ps}

\figcaption{Spectra of 14 major zones of the M86, in correspondence with Tab.\ref{t:mo}. Spectra are assembled in rows with increasing number from left to right and from up to down. 
\label{f:speall}
}

\end{figure*}

Once the detailed spectroscopic analysis has been performed, we could try to
simplify the results by putting together all the regions of similar
temperature. The minimum width of the region in this analysis is one
arcminute, which also reduces possible influence of the point spread
function on the analysis. Such region selection allows us to tabulate the
basic properties of the M86 ISM as well as take advantage of good statistics
to look whether the spectral model adopted for the refined analysis is still
a valid one. Instead of fitting the element abundance of all the elements,
we use the photospheric solar abundance pattern of (Anders \& Grevesse 1989)
and in Fig.\ref{f:speall} we demonstrate the quality of the fit. It becomes
clear from this comparison that O is underabundant, while high energy lines,
especially Si He-like triplet, is located on the blue side of the model,
possibly implying an overcorrection for the pn CTI losses. Other than that,
the adopted model provides a good description of the spectrum with
parameters listed in Tab.\ref{t:mo}. Column (1) presents the region label,
as indicated in Fig.\ref{imh}, (2) lists the name of the region, (3--4)
temperature of two thermal components, (5) element abundance relative to the
solar pattern of Anders \& Grevesse (1989), (6--7) normalizations of the two
components in XSPEC units, (8--9) reduced $\chi^2$ and a number of degrees
of freedom, which reach the finest possible energy binning for most
spectra. Columns (10--12) present the entropy, pressure and gas mass of a
sum of two components, assuming they occupy the same volume.  Uncertainties
in the parameter estimates are cited at the 68\% confidence level and do not
include any systematical uncertainties, either instrumental or model.

The mass-averaged properties of M86 within the central 80 kpc are
$kT=1.17\pm0.02$ keV, $Fe=0.47\pm0.01Fe_\odot$, $S=78\pm2$ keV cm$^2$,
$P=(3.73\pm0.05)\times 10^{-12}$ ergs cm$^{-3}$. The total gas mass is
$(10.18\pm0.11)\times 10^{10} M_\odot$. The central 30 kpc of M86 are
characterized by $kT=1.08\pm0.03$, $Fe=0.95\pm0.03Fe_\odot$, $S=55\pm2$ keV
cm$^2$, $P=5.33\pm0.11 10^{-12}$ ergs cm$^{-3}$, mass-averaged over the total
gas mass of $(2.03\pm0.04)\times 10^{10} M_\odot$.

\begin{table*}[ht]
{
\begin{center}
\footnotesize
{\renewcommand{\arraystretch}{0.9}\renewcommand{\tabcolsep}{0.12cm}
\caption{\footnotesize
Properties of main regions of M86. 
\label{t:mo}}

\begin{tabular}{cccccccccccc}
 \hline
N & Name &$kT_1$ & $kT_2$ & $Z/Z_\odot$ & $norm_1$ & $norm_2$ & $\chi^2$ & N & S & P, $10^{-12}$ & $M_{\rm gas}$\\
  &  & keV  & keV     &              & $10^{-3}$ & $10^{-3}$ & & d.o.f. & keV cm$^2$ & ergs cm$^{-3}$ & $10^{10} M_\odot$  \\
\hline
01 & NE arm&$0.690\pm0.013$ & $1.337\pm0.061$ & $0.64\pm0.04$ & $0.79\pm0.10$ & $0.41\pm0.07$ &1.17 & 213&$42\pm4$&  $ 5.3\pm  0.3$&  $0.31\pm  0.02$\\
02 & core&$0.677\pm0.007$ & $1.188\pm0.047$ & $0.76\pm0.05$ & $0.41\pm0.05$ & $0.19\pm0.03$ &1.31& 171   &$14\pm1$&  $21.4\pm  1.1$&  $0.03\pm  0.01$\\
03 & SE ext.&$0.659\pm0.020$ & $1.027\pm0.066$ & $0.56\pm0.05$ & $0.23\pm0.06$ & $0.14\pm0.05$ &1.24& 115&$16\pm3$&  $15.1\pm  1.7$&  $0.03\pm  0.01$\\
04 & shock: W&$0.831\pm0.005$ & $1.369\pm0.036$ & $0.98\pm0.04$ & $1.21\pm0.09$ & $1.40\pm0.10$ &1.63&331&$63\pm 3$& $ 4.0\pm  0.1$&  $1.04\pm  0.03$\\
05 & east-1&$0.741\pm0.018$ & $1.386\pm0.222$ & $0.29\pm0.03$ & $0.67\pm0.13$ & $0.44\pm0.12$ &0.92 & 120&$58\pm10$& $ 3.8\pm  0.5$&  $0.44\pm  0.03$\\
06 & shock: E&$1.605\pm0.141$ & $0.774\pm0.017$ & $0.67\pm0.09$ & $0.19\pm0.04$ & $0.17\pm0.04$ &0.97 &74&$56\pm 9$& $ 6.0\pm  0.6$&  $0.11\pm  0.01$\\
07 & plume&$0.792\pm0.004$ & $1.566\pm0.079$ & $1.18\pm0.06$ & $1.12\pm0.10$ & $0.44\pm0.06$ &2.15& 282  &$51\pm 4$& $ 5.3\pm  0.2$&  $0.44\pm  0.02$\\
08 & trans.&$0.726\pm0.022$ & $1.278\pm0.073$ & $0.88\pm0.07$ & $0.29\pm0.06$ & $0.21\pm0.05$ &1.16& 163 &$26\pm 4$& $10.8\pm  0.9$&  $0.07\pm  0.01$\\
09 &east-3& $1.247\pm0.097$ & $0.666\pm0.062$ & $0.10\pm0.02$ & $1.04\pm0.20$ & $0.47\pm0.18$ &1.14& 101 &$59\pm10$& $ 3.5\pm  0.3$&  $0.61\pm  0.06$\\
10 &east-2& $1.333\pm0.041$ & $0.666\pm0.020$ & $0.29\pm0.02$ & $1.21\pm0.14$ & $0.72\pm0.12$ &1.10& 199 &$65\pm 6$& $ 3.3\pm  0.1$&  $0.86\pm  0.04$\\
11 & trans.&$0.811\pm0.041$ & $1.369\pm0.192$ & $0.36\pm0.06$ & $0.10\pm0.04$ & $0.17\pm0.04$ &1.10& 48  &$53\pm12$& $ 5.5\pm  0.7$&  $0.08\pm  0.01$\\
12 & trans.&$0.815\pm0.020$ & $1.512\pm0.146$ & $0.59\pm0.08$ & $0.14\pm0.04$ & $0.23\pm0.05$ &1.48& 63  &$56\pm 9$& $ 6.0\pm  0.6$&  $0.11\pm  0.01$\\
13 & group SW&$1.483\pm0.027$ & $0.846\pm0.006$ & $0.44\pm0.01$ & $7.49\pm0.17$ & $2.30\pm0.16$ &1.36&508&$91\pm 2$& $ 3.2\pm  0.1$&  $5.13\pm  0.07$\\
14 & Void&$1.234\pm0.022$ & --- & $0.12\pm0.01$ & $2.82\pm0.14$ & --- &1.15 & 242                        &$98\pm 5$& $ 2.7\pm  0.1$&  $0.92\pm  0.02$\\
\hline
\end{tabular}
}
\end{center}
}
\vspace*{-1.cm}
\end{table*}

\section{Discussion}\label{s:disc}

XMM-Newton observations presented here, while confirming previous arguments
in favor of strong interactions taking place in M86, provide new insights as
to their yet unclear nature.  As already mentioned in the introduction
previous works concentrated on M86 galaxy interaction with Virgo's
ICM. However, a number of observations challenge this picture:
the distance of $2.4\pm1.4$ Mpc between the M86 and M87 (Neilsen \&
Tsvetanov 2000).  This result implies that M86 lies outside the virial
radius of the Virgo cluster.  Also, based on the RASS surface brightness
measurements (B\"ohringer et al. 1994) the X-ray emission surrounding the
M86 galaxy is most naturally associated to M86's group medium, as opposed to
Virgo ICM as often assumed in previous studies. XMM-Newton provides an
additional proof to this picture by measuring the 1.2 keV temperature of the
group component, compared to 2.4 keV temperature expected from the Virgo
cluster.

Although we dismiss Virgo ICM (more correctly, M87 cloud) as responsible for
M86 X-ray morphological disruption, obviously our depiction still requires
the presence of a dense medium for M86 to interact with.  We assume for the
sake of simplicity that M86 velocity with respect to the putative medium is
the same as that relative to M87, that is about 1500 km s$^{-1}$.  Then we
find that a medium number density no less than a few $\times 10^{-4}$ is
required in order for the observed post-shock pressure of order
$5\times10^{-12}$ ergs cm$^{-3}$ to be generated.  Such diffuse gas should
be detectable in X-ray emission if it had a temperature in excess of 0.2
keV. In fact, B\"ohringer et al. (1994, see their Fig.2) had already pointed
out the presence of a large-scale X-ray filament in precisely that region. It
was revealed in the RASS data as residual emission after removal of the
contributions associated with both M87 and M86. The density of such
structure was typically found about $10^{-3}$ cm$^{-3}$, sufficiently high
to justify the interpretation presented above on M86 dynamics.  Its width
(on the plane of the sky) is about 300 kpc, typical for a filament, yet the
high density suggests a high degree of virialization such as that
characteristic of a chain of groups. Fig.\ref{f:vcc} summarizes the
proposed interaction geometry.
We note that both the X-ray emission and
infalling galaxies are present at the position of M86. It is remarkable that
the alignment of the residual X-ray emission is parallel to the suggested
direction of motion of M86 in the observer plane to the south-west.

\includegraphics[width=8.4cm]{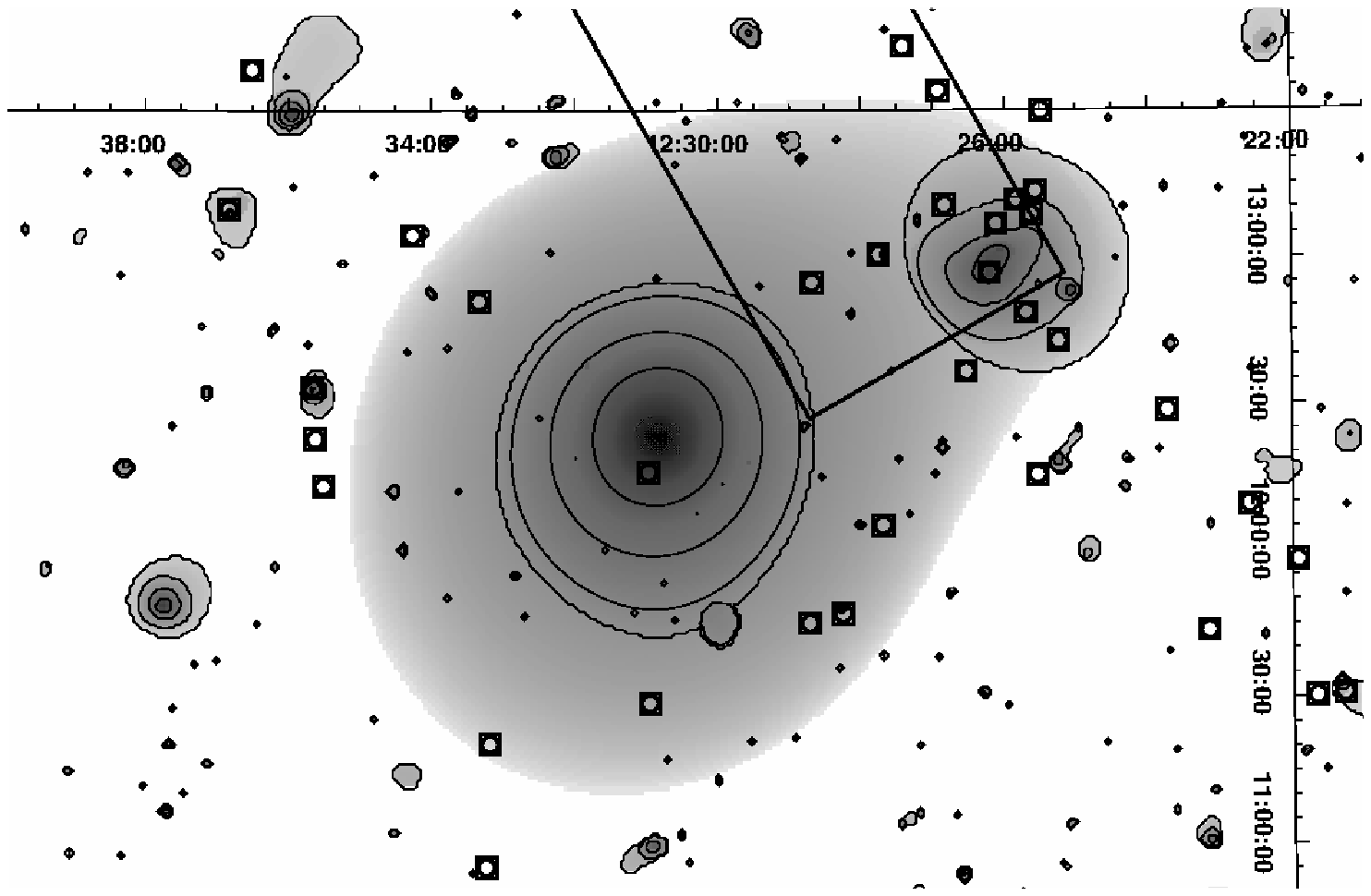}

\figcaption{Large-scale structure around M86. The wavelet-reconstructed
image reproduces a part of ROSAT All-Sky Survey in the 0.5--2 keV energy
band, close to M86. Black squares indicate the galaxies falling onto Virgo
from behind ($V_{M87}-V_{galaxy} > 600$ km/s), using the galaxy velocity
data from Bingelli et al. (1985; 1993). Large rectangular region denotes the
position of the excess X-ray emission identified in B\"ohringer et
al. (1994).
\label{f:vcc}
}

Presence of low-entropy tails behind the M86 core, such as plume and
northeastern arm strongly suggest that M86 has already been {\it harassed},
prior to the generation current shock wave.  Quite strikingly the
large-scale emission morphology of M86 group displayed in Fig.\ref{f:vcc}
appears relaxed, while strong interactions are taking place in the core of
M86 galaxy. This suggests that the interactions responsible for the
morphological disturbances in M86 galaxy were produced by objects which did
not affect significantly the gas in M86 groups. One such possibility is
provided by high velocity galaxy-galaxy encounters. Therefore, M86 system
could be provide an important case for establishing the morphology-density
relation between one and two virial radii, as suggested by the SDSS results
(e.g. Goto et al. 2003).

The possibility of galaxy-galaxy interactions prior to infall onto a cluster
is not remote.  M86 group appears to be attracted toward the Virgo
supercluster with mass roughly $\sim10^{15} M_\odot$ (see also Neilsen \&
Tsvetanov 2000).  Numerous other galaxies accreting from different distances
will have different accretion velocities as well as directions allowing for
a high-speed interactions to take place. Most observers agree that the
direction of the motion of M86 in the observer's plane is towards the south
with some component to the west. So, M86 is aiming to the point of
interaction of M87 and M49, but slightly bears off west. Another dozen of
infalling galaxies, displayed in Fig.\ref{f:vcc} seems to share the same
route and some of them could be the former collision companions to M86.  The
magnitude of the interaction, recorded in M86 X-ray image, should have had a
strong effect on those galaxies, leading to substantial depletion in the HI
content. Our estimate of the ram pressure to be of the order of
$4\times10^{-12}$ ergs cm$^{-3}$ is similar to ram pressure expected to act
on galaxies within the Virgo cluster, as described in Vollmer et al. (2001).

\section{Conclusions}\label{s:conc}

An in-depth understanding of the processes in the hot interstellar medium of
M86, provided by observations of M86 has suggested a number of past and
on-going interactions to take place. Comparison with the position of M86 in
the Virgo cluster, promotes the environment outside the virial radius of M87
cloud to bear the prime responsibility for the apparent morphological
transformation of the X-ray appearance of M86. Some of the past interactions
in M86 are characterized by 10--50 kpc scale, which is only suitable for
galaxy-galaxy interactions on one hand, setting limits on a degree of
interaction, on the other, thus providing us with understanding of the
processes leading to establishment of the morphology-density relations in
clusters of galaxies.

\smallskip

{\it Acknowledgments.} This paper is based on observations obtained
with XMM-Newton, an ESA science mission with instruments and
contributions directly funded by ESA Member States and the USA
(NASA). The XMM-Newton project is supported by the Bundesministerium
f\"{u}r Bildung und Forschung/Deutsches Zentrum f\"{u}r Luft- und
Raumfahrt (BMFT/DLR), the Max-Planck Society and the
Heidenhain-Stiftung, and also by PPARC, CEA, CNES, and ASI. AF
acknowledges receiving the Max-Plank-Gesellschaft Fellowship.  FM was
partially supported by the European Community through the Research and
Training Network - `The Physics of the Intergalactic Medium' - under
contract HPRN-CT2000-00126 RG29185. AF acknowledges useful discussions
with Hans B\"ohringer, Bernd Vollmer, Jacqueline van Gorkom and
Cristina Popescu. 
The authors thank the anonymous referee for useful criticism. 

\vspace*{-0.5cm}

\bibliographystyle{aabib99}

\end{document}